\documentclass[12pt]{iopart}
\expandafter\let\csname equation*\endcsname\relax
\expandafter\let\csname endequation*\endcsname\relax 
\usepackage{dcolumn}
\usepackage{bm}
\usepackage{graphics}
\usepackage{longtable}
\usepackage{units}
\usepackage{textcomp}
\usepackage{float}
\usepackage{mathtools}
\usepackage{amsfonts}
\usepackage[font=footnotesize]{caption}

\begin{document}

\title{Vibrational dephasing in matter-wave interferometers}

\author{A Rembold$^1$, G Sch\"{u}tz$^1$, R R\"{o}pke$^1$, W T Chang$^2$, I S Hwang$^2$, A G\"{u}nther$^{3,*}$ and A Stibor$^{1,+}$}
\address{$^1$ Institute of Physics and Center for Collective Quantum Phenomena in LISA$^+$,
University of T\"{u}bingen, Auf der Morgenstelle 15, 72076 T\"{u}bingen, Germany}
\address{$^2$ Institute of Physics, Academia Sinica, Nankang, Taipei 11529, Taiwan, Republic of China}
\address{$^3$ Institute of Physics and Center for Collective Quantum Phenomena in LISA$^+$,
University of T\"{u}bingen, Auf der Morgenstelle 14, 72076 T\"{u}bingen, Germany}
\ead{$^*$a.guenther@uni-tuebingen.de and $^+$alexander.stibor@uni-tuebingen.de}

\begin{abstract}
Matter-wave interferometry is a highly sensitive tool to measure small perturbations in a quantum system. This property allows the creation of precision sensors for dephasing mechanisms such as mechanical vibrations. They are a challenge for phase measurements under perturbing conditions that cannot be perfectly decoupled from the interferometer, e.g.~for mobile interferometric devices or vibrations with a broad frequency range. Here, we demonstrate a method based on second-order correlation theory in combination with Fourier analysis, to use an electron interferometer as a sensor that precisely characterizes the mechanical vibration spectrum of the interferometer. Using the high spatial and temporal single-particle resolution of a delay line detector, the data allows to reveal the original contrast and spatial periodicity of the interference pattern from ``washed-out" matter-wave interferograms that have been vibrationally disturbed in the frequency region between 100 and \unit[1000]{Hz}. Other than with electromagnetic dephasing, due to excitations of higher harmonics and additional frequencies induced from the environment, the parts in the setup oscillate with frequencies that can be different to the applied ones. The developed numerical search algorithm is capable to determine those unknown oscillations and corresponding amplitudes. The technique can identify vibrational dephasing and decrease damping and shielding requirements in electron, ion, neutron, atom and molecule interferometers that generate a spatial fringe pattern on the detector plane.
\end{abstract}

\section{Introduction}
The high phase sensitivity of interferometric sensors is the basis for their broad implementation in technical \cite{Grattan2013} as well as in fundamental applications \cite{Abbott2016,Graham2013}. Recent developments in matter-wave interferometry indicate the wide applicability in various fields of quantum physics. Such interferometers are used for interferometry with large organic molecules \cite{Gerlich2011}, to test the limits of quantum mechanical superpositions \cite{Arndt2014a}, for interference on optical ionization gratings in the time domain \cite{Haslinger2013}, for the measurement of inertial forces \cite{gustavson1997,Hasselbach1993}, to determine gravitational acceleration \cite{Peters1999} or with coherent particles prepared as self-interfering clocks \cite{Margalit2015,Arndt2015a}.

Interferometers are also highly sensitive towards mechanical vibrations \cite{Stibor2005}. Such perturbations dephase the matter-wave and decrease the interference contrast. This is in particular critical for precision experiments in a perturbing environment. Vibrational dephasing has been analyzed and decreased in several related fields of research e.g.~for a continuous beam of thermal atoms in a Mach-Zehnder interferometer \cite{Miffre2006}. In precision interferometric measurements with ultracold atoms, such as for gravity \cite{Hauth2013} or inertial effects \cite{Geiger2011}, the vibrations induce an arbitrary phase shift for each interfering particle pulse. Usually, the atoms have a large time of flight in the setup and therefore an elaborated active and passive vibration damping is required \cite{Hensley1999}. The vibrational phase shifts can be compensated e.g.~with the signal of a low noise seismometer attached to the beam mirror \cite{leGouet2008} or by simultaneous operation of a pair of conjugate atom interferometers \cite{Chiow2009,Chiow2011}. Vibrational noise is also one of the main challenges to achieve a higher accuracy in measurements concerning the equivalence principle \cite{Fray2004,Chen2014} since it is impossible to distinguish between the gravitational acceleration and a perturbing movement of the setup. For precise tests of the weak equivalence principle, the differential phase between dual-species atom interferometers can be extracted using a mechanical accelerometer to measure the vibration-induced phase and to reconstruct the interference contrast \cite{Barrett2015}. Dephasing noise reduction has also applications to increase current frequency standards for atomic clocks. By a phase lock of a classical oscillator to an atomic superposition state, based on repeated coherence-preserving measurements and phase corrections, an atomic clock can be operated beyond the limit set by the local oscillator noise \cite{Kohlhaas2015}. Furthermore, the laser noise in the stability between different optical clocks can be decreased, allowing probe times longer than the coherence of the laser in the time domain \cite{Hume2016}.

It is also important to identify dephasing in experiments to study the theory of decoherence \cite{Zurek2003,Pikovski2015}. Thereby, the gradual loss of interference contrast due to entanglement of the quantum superposition state of the matter-wave with the environment is measured \cite{Hackermuller2004,Hornberger2003,Sonnentag2007} and needs to be distinguished from the contrast loss originating from dephasing. Such mechanisms also significantly disturb sensitive phase measurements, as necessary in Aharonov-Bohm physics \cite{Aharonov1959,Batelaan2009,Schmid1985,Schuetz2015b}. 

Recently, we demonstrated in theory and experiment a method to reastablish an electron interference pattern disturbed by known single \cite{Rembold2014} and multifrequency \cite{Guenther2015}  electromagnetic perturbations using second-order correlation analysis \cite{Folling2005}. Thereby, the technique is based on the high spatial and temporal resolution of a delay line detector \cite{Jagutzki2002} for single-particle events. In this article, we demonstrate that this method can be applied on vibrational dephasing and be extended to perform spectroscopy of unknown dephasing perturbation frequencies from a ``washed-out" interference pattern. We present a precise characterization of the mechanical resonance spectrum of our electron interferometer after applying vibrational dephasing perturbations. In contrast to previous measurements with electromagnetic disturbances, the actual perturbation frequencies can vary from the applied ones due to possible excitations of higher harmonics in the setup and additional perturbation frequencies originating from the environment. Therefore, a numerical search algorithm has been developed to identify the unknown perturbation frequencies and corresponding amplitudes. As for electromagnetic oscillations, it is again possible to fully correct the dephasing by our correlation method, reestablishing interferograms with a high contrast. Furthermore, the influence of temporal binning of the measurement data is analysed in detail.

The mechanical dephasing perturbations were artificially applied in a biprism electron interferometer by a speaker and piezo element in a frequency range between \unit[100 and 1000]{Hz}. This kind of frequencies occur in typical lab situations when acoustic noise, vibrations from the building, the cooling system or the vacuum pumps decrease the interference contrast and therefore ``wash-out" the matter-wave interferogram. Because of the complexity of the system with several mechanical resonances, the contributing perturbation frequencies are not known a priori. Therefore, our numerical search algorithm was developed, combining the second-order correlation theory with a Fourier analysis. According to the Wiener-Khintchine theorem \cite{Wiener1930,Khintchine1934}, the Fourier transform of the correlation function equals the power spectrum of the perturbed measurement signal. This is used in our method to identify the perturbation frequencies and amplitudes that have contributed to the dephasing of the interference pattern. With these values it is possible to reconstruct the original undisturbed pattern. The contrast of the unperturbed interference pattern could be recovered in the whole frequency range. Our technique allows to reveal the matter-wave nature of particles under conditions in which usual spatial integration of an interference pattern would be inapplicable.

The method has potential applications to restore the contrast for interferometers in perturbing environments that cannot be satisfactorily decoupled by damping or shielding in a broad frequency range. In case, the time that the particles need to cross the interferometer is significantly smaller than the cycle duration of the perturbation, our technique can reveal the spectrum of vibrational and electromagnetic frequencies and amplitudes in all interferometers that generate a spatial fringe pattern on a detector with a high spatial and temporal single-particle resolution. Such detectors exist for electrons \cite{Jagutzki2002}, ions \cite{Jagutzki2002}, neutrons \cite{Siegmund2007}, atoms \cite{Schellekens2005} and molecules \cite{Zhou2012}. The technique is therefore also a helpful tool to design optimal active and passive damping structures for a specific setup.

\section{Theory}
To calculate the two-dimensional second-order correlation function $g^{(2)}(u,\tau)$, we start with the probability distribution $f(y,t)$ of the particle impacts at the detector
\begin{equation}
f(y,t) = f_0\Big(1+K_0\cos \big(ky + \varphi\left(t\right)\big)\Big) ~, 
\label{eq1} 
\end{equation}
where $K_0$ is the unperturbed contrast, $k=2\pi/\lambda$ the wave number of the unperturbed interference pattern, with $\lambda$ the spatial periodicity, and the normalization factor $f_0$. The interference pattern is perturbed by the time-dependent phase shift $\varphi\left(t\right)$, which consists of a superposition of $N$ frequencies $\omega_j$
\begin{equation}
\varphi(t)=\sum_{j=1}^N \varphi_j\cos\left(\omega_j t + \phi_j\right) ~. 
\label{eq2}
\end{equation}
Here $\varphi_j$ and $\phi_j$ denote the peak phase deviation and phase of the perturbation frequency $\omega_j$ respectively. This phase shift leads to a washout of the integrated interference pattern at the detector \cite{Rembold2014,Guenther2015}, yielding
\begin{align}\label{eq3}
\lim_{T\rightarrow\infty} \frac{1}{T}\int_0^T f(y,t)\, \mathrm{d}t &= f_0+\lim_{T\rightarrow\infty}\frac{f_0 K_0}{2T}\left(\mbox{e}^{iky}\int_0^T \mbox{e}^{i\varphi_1\cos(\omega_1 t+\phi_1)}\,\mathrm{d}t+c.c.\right) \\ \nonumber
&= f_0\big(1+K_0\,\!J_0(\varphi_1)\cos(ky)\big)
\end{align}
for $N=1$. Here, $\mbox{e}^{\pm i\varphi_1\cos(\omega_1 t+\phi_1)} = \sum_{n_1=-\infty}^{+\infty}\, \!J_{n_1}(\varphi_1)\mbox{e}^{in_1\left(\omega_1 t+\phi_1\pm\frac{\pi}{2}\right)}$ was used. The limit of the time integral is equal to one only for $n_1=0$ and zero otherwise. The contrast in the ``washed-out" interference pattern is thus reduced by a factor of $J_0(\varphi_1)\approx 1-\varphi_1^2/4$ for small peak phase deviations $\varphi_1<1$. 

According to \cite{Rembold2014,Guenther2015} the second-order correlation function is calculated and the explicit correlation function for $N$ perturbation frequencies $\omega_j$ becomes
\begin{equation} \label{eq4}
g^{(2)}(u,\tau) = 1 + \frac{1}{2}K_0^2\sum_{\substack{\left\{n_j, m_j\right\}\in\mathbb{Z}\\ j=1\ldots N}}A_{\{n_j, m_j\}}\left(\tau,\Phi_{\{n_j, m_j\}}\right)\cos\left(ku+\tilde{\varphi}_{\{n_j, m_j\}}\right) ~,
\end{equation}
with
\begin{equation} \label{eq5}
A_{\{n_j, m_j\}}\left(\tau,\Phi_{\{n_j, m_j\}}\right) = \tilde{B}_{\{n_j, m_j\}}\left(\varphi_j\right)\cdot\cos\Biggl(\sum_{j=1}^N m_j\omega_j\tau+\Phi_{\{n_j, m_j\}}\Biggr)~.
\end{equation}
The sum in equation (\ref{eq4}) has to be taken over all integer multiplets $\left\{n_j, m_j\right\}\in \mathbb{Z}~,~ j=1\ldots N$, for which the following constraint is fulfilled \cite{Guenther2015}
\begin{equation}
\sum_{j=1}^N\left(n_j+m_j\right)\omega_j=0 ~.
\label{eq6}
\end{equation}
For a finite acquisition time $T$, the constraint in equation (\ref{eq6}) has to be modified to $\left|\sum_{j=1}^N\left(n_j+m_j\right)\omega_j\right|< 2\pi/T$, as $1/T$ defines the minimal resolvable frequency. In principle an infinite number of integer multiplets fulfil this constraint, but the contribution to the sum is suppressed due to the strong decay of the Bessel functions of first kind $\!J_n$ in
\begin{equation} \label{eq7}
\tilde{B}_{\{n_j, m_j\}}\left(\varphi_j\right)\coloneqq \prod_{j=1}^N J_{n_j}\left(\varphi_j\right)J_{m_j}\left(\varphi_j\right) ~.
\end{equation}
The spatial correlation phase $\tilde{\varphi}_{\{n_j, m_j\}}$ and temporal phase $\Phi_{\{n_j, m_j\}}$ in equation (\ref{eq4}) and (\ref{eq5}) are given by
\begin{align} \label{eq8}
\tilde{\varphi}_{\{n_j, m_j\}} &\coloneqq \frac{\pi}{2}\sum_{j=1}^N\left(m_j-n_j\right) ~, \\ \nonumber 
\Phi_{\{n_j, m_j\}} &\coloneqq \sum_{j=1}^N \phi_j\left(m_j+n_j\right)~.
\end{align}

If the constraint in equation (\ref{eq6}) is satisfied only for $n_j=-m_j$ $(j=1\ldots N)$, the temporal phase $\Phi_{\{-m_j, m_j\}}$ in equation (\ref{eq8}) becomes zero and $\tilde{\varphi}_{\{-m_j,m_j\}} =\pi\sum_{j=1}^N m_j$. Together with $\!J_{-m_j}(\varphi_j)=\left(-1\right)^{m_j}\!J_{m_j}(\varphi_j)$ equation (\ref{eq4}) then simplifies, yielding the approximate correlation function \cite{Guenther2015} 
\begin{equation}
g^{(2)}(u,\tau) = 1 + A\left(\tau\right)\cos\left(ku\right) ~,\label{eq9}
\end{equation}
with
\begin{equation}
A(\tau) = \frac{1}{2}K_0^2\prod_{j=1}^N \sum_{m_j=-\infty}^\infty J_{m_j}\left(\varphi_j\right)^2\cos\left(m_j\omega_j\tau\right) ~.\label{eq10}
\end{equation}
More details to the differentiation and applicability of the explicit and approximate correlation theory can be found in \cite{Guenther2015}.

The correlation functions in equation (\ref{eq4}) and (\ref{eq9}) show a periodic modulation in the spatial distance $u$, with the same periodicity $\lambda$ as in the unperturbed interference pattern. The amplitudes $A_{\{n_j, m_j\}}\left(\tau,\Phi_{\{n_j, m_j\}}\right)$ and $A(\tau)$ of this modulation result from the specific perturbation spectrum. The maximum of $0.5 K_0^2$ is achieved at $\tau = 0$, where only the addends with $n_j=-m_j$ contribute to the correlation function and therefore equation (\ref{eq4}) is equal to equation (\ref{eq9}), resulting in \cite{Rembold2014,Guenther2015}
\begin{equation}
g^{(2)}(u,0) = 1 + \frac{1}{2}K_0^2\cos\left(ku\right) ~,\label{eq11}
\end{equation} 
which is appropriate to obtain the contrast of the unperturbed interference pattern $K_0$ and the spatial periodicity $\lambda=2\pi/k$. Therefore, it is possible to proof matter-wave interference, although the periodic pattern would be ``washed-out" after integration of the signal.

The function $A_{\{n_j, m_j\}}\left(\tau,\Phi_{\{n_j, m_j\}}\right)$ in equation (\ref{eq5}) contains a superposition of harmonics, intermodulation terms (sums and differences) of the perturbation frequencies at discrete values of $\sum_{j=1}^N m_j\omega_j$, with the coefficient $m_j\in\mathbb{Z}$ resulting from the constraint in equation (\ref{eq6}). Their amplitudes are given by the peak phase deviations $\varphi_j$ via the product of the Bessel functions in equation (\ref{eq7}). The highest contributing frequency component appears at roughly $\sum_{j=1}^N m_{j,max} \omega_j$, with $m_{j,max}\approx\varphi_j$. For higher orders $m_j>\varphi_j$, the Bessel function decays strongly and the frequency component disappears. To get the positions of the frequencies and the corresponding amplitudes, the amplitude spectrum $\left|\mathcal{F}\left(g^{(2)}(u,\tau)\right)(u,\omega)\right|$ of equation (\ref{eq4}) for the positive frequency region is calculated to
\begin{align} \label{eq12}
\frac{1}{2\pi}\left|\mathcal{F}\left(g^{(2)}(u,\tau)\right)(u,\omega)\right|^2 = &\,  \delta(\omega)^2+2\cdot\Bigg[\frac{1}{2}K_0^2\sum\limits_{\substack{\left\{n_j, m_j\right\}\in\mathbb{Z}\\ j=1\ldots N}} \tilde{B}_{\{n_j, m_j\}}\left(\varphi_j\right)\cdot\\ \nonumber 
&\cdot\delta\left(\omega-\omega_{\{m_j\}}\right)\cdot\cos\left(\Phi_{\{n_j, m_j\}}+\frac{\pi}{4}\right)\cos\left(ku+\tilde{\varphi}_{\{n_j, m_j\}}\right)\Bigg]^2 ~, \nonumber
\end{align}
with the Dirac delta function $\delta\left(\omega\right)$ and the frequency components $\omega_{\{m_j\}} \coloneqq \sum\limits_{j=1}^{N}m_j\omega_j$.

The amplitude spectrum of the correlation function is used to identify the perturbation frequencies $\omega_j$, peak phase deviations $\varphi_j$ and phases $\phi_j$ of the applied perturbation.

\section{Experiment} \label{Experiment}

\begin{figure}
\centering
\includegraphics[width=0.6\textwidth]{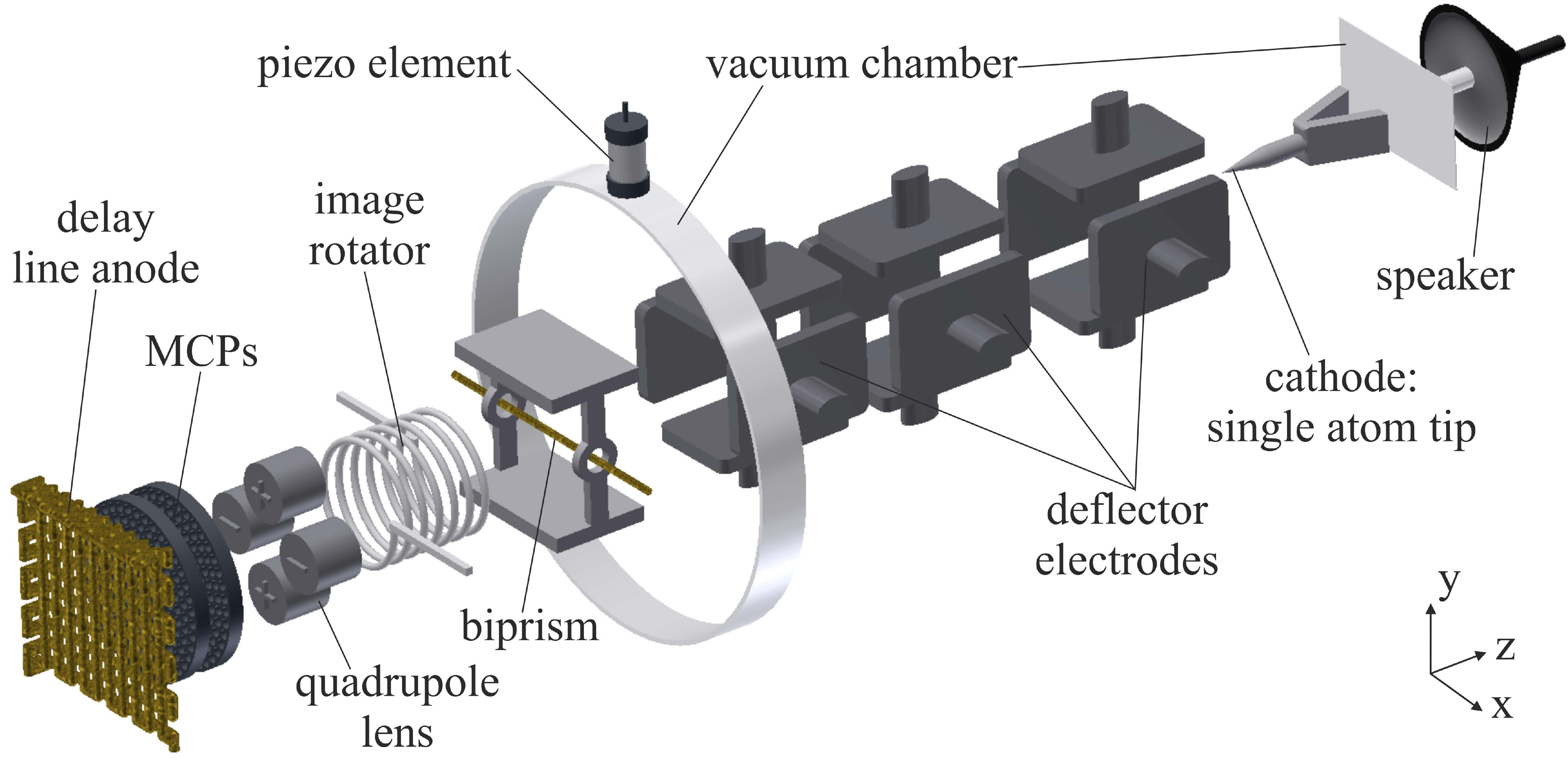} \caption{(Color online) Sketch of the experimental setup for the measurement of vibrational dephasing in an electron biprism interferometer. An electron beam is emitted by a single atom tip cathode and aligned by deflection electrodes. After separation by a charged biprism fiber, the partial matter waves are superposed and form an interference pattern that can be tilted by an image rotator parallel to the $x$-direction of the quadrupole lens. The magnified pattern is detected by two multi-channelplates (MCPs) in combination with a delay line anode. To demonstrate the technique for identifying perturbation frequencies of mechanical vibrations and correcting the reduction of contrast, vibrations were artificially introduced by a speaker and a piezo element mounted outside on the vacuum chamber.}
\label{fig1}
\end{figure}

We demonstrate the identification of vibrational dephasing in an electron biprism interferometer \cite{Mollenstedt1956a}. The experimental setup is illustrated in figure \ref{fig1} and described elsewhere \cite{Schuetz2014,Hasselbach2010,Hasselbach1998a,Maier1997}. A coherent electron beam is field emitted by a single atom tip source \cite{Kuo2006a,Kuo2008}. The beam is adjusted by electrostatic deflection electrodes towards a \unit[400]{nm-thick} biprism fiber that is coated with gold-palladium \cite{Schuetz2014}. It is positioned between two grounded electrodes and acts as a coherent beam splitter for the electron matter-wave \cite{Mollenstedt1956a}. By the application of a positive voltage the two separated beam paths get deflected towards each other creating a matter-wave interference pattern with a period of a few hundred nanometers. The quadrupole lens expands the pattern by a factor of several thousand which is then projected on a delay line detector. Using the image rotator the pattern is aligned parallel to the $x$-direction. The detector amplifies the single electron events by two multi-channelplates and detects them with high spatial and temporal resolution \cite{Jagutzki2002}. The individual components are mounted on two ceramic rods to prevent temperature drifts and provide electrical insulation. The whole system is installed within an ultrahigh vacuum chamber at a pressure of \unit[$4.5\times10^{-10}$]{mbar}. To avoid perturbations it is mounted on an air-damped optical table and shielded against electromagnetic radiation.

For the demonstration of dephasing identification and frequency analysis, the electron interferences are artificially disturbed by mechanical vibrations from a speaker in the frequency range between \unit[100 and 320]{Hz} and a piezo element in the range between \unit[330 and 1000]{Hz}. Both are mounted outside on the vacuum-chamber. The speaker was attached behind the cathode generating vibrations along the $z$-direction (see figure~\ref{fig1}), with a direct connection to the vacuum chamber. The piezo element mounted on the vacuum chamber produced vibrations in the $y$-direction. Both are controlled by a frequency generator with a resolution of \unit[1]{\textmu Hz} and an accuracy of $\unit[\pm10]{ppm}$ on the set frequency value.

Only a single excitation frequency is applied at once. At each frequency an interference pattern with $(1.95\pm0.02)\times10^5$ electrons (at a count rate of \unit[$(2.0\pm0.5)$]{kHz} for the speaker measurement and \unit[$(11\pm2)$]{kHz} for the piezo measurement) is accumulated and the temporal and spatial information for each particle is recorded. Stepwise, the frequency is increased and a new interference pattern is acquired. This way, the complete spectral response of the interferometer was measured. The electron energy for each measurement was \unit[1.45]{keV} which results in a velocity of \unit[$2.26\times 10^7$]{m/s}. The flight time of the electrons from the tip to the delay line detector amounts \unit[26]{ns}.

\section{Data Analysis} \label{Data}
We will demonstrate exemplarily our method to analyse an electron interference pattern perturbed by a mechanical vibration with the excitation frequency of $\omega_0/2\pi = \unit[540]{Hz}$, which is introduced by a piezo element. 

\begin{figure}
\centering
\includegraphics[width=0.6\textwidth]{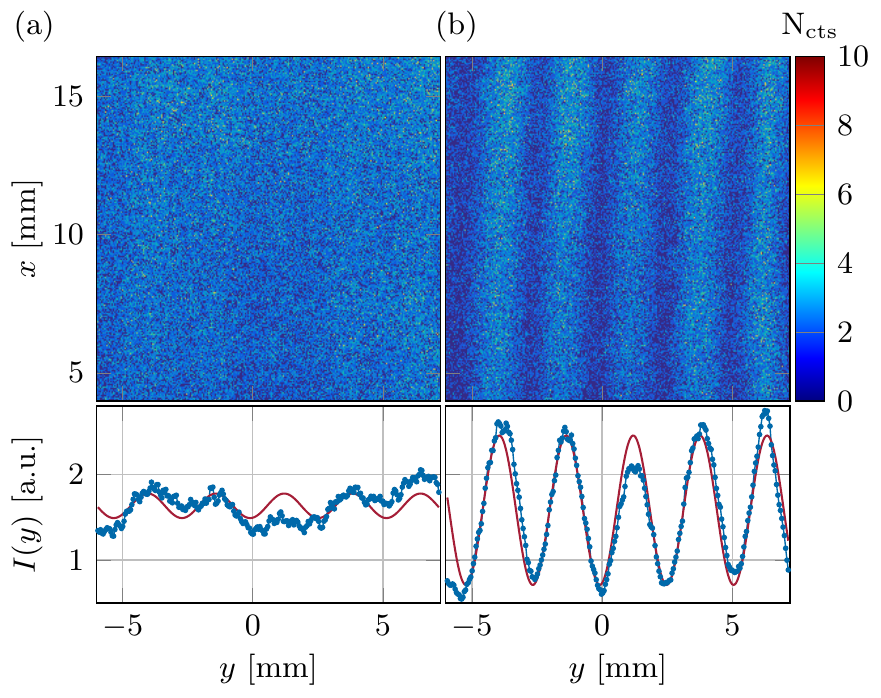} \caption{(Color online) (a) Electron interference pattern perturbed with an excitation frequency of $\omega_0/2\pi = \unit[540]{Hz}$. The averaged interference pattern (blue dots) and the fitted model function in equation (\ref{eq13}) (red solid line) are presented below. (b) With the obtained parameters from the correlation analysis and subsequent optimization the unperturbed interference pattern can be reconstructed.}
\label{fig2}
\end{figure}

For each electron that reaches the detector the spatial positions ($x_i$, $y_i$) and the arrival time $t_i$ is recorded. The histogram for the integrated signal is shown in figure \ref{fig2}(a). To determine the contrast of the perturbed interference pattern $K_{\text{pert}}$ and spatial periodicity $\lambda_{\text{pert}}$, the histogram is averaged along the $x$-direction and fitted with a model function
\begin{equation}
I(y) = I_0\big(1+K\cos(ky)\big) ~,
\label{eq13}
\end{equation}
with the mean intensity $I_0$, contrast $K$ and wave number $k=2\pi/\lambda$. The result is plotted below the histogram in figure \ref{fig2}(a), yielding a contrast $K_{\text{pert}}=\unit[(8.8\pm 1.6)]{\%}$ and a spatial periodicity $\lambda_{\text{pert}}= \unit[(2.62\pm0.06)]{mm}$. Here, the errors indicate the $\unit[95]{\%}$ confidence interval of the fit.

To extract the two-dimensional correlation function $g^{(2)}(u,\tau)$, a histogram $N_{u,\tau}$ of all particle pair distances $(y_i-y_j)$ and time differences $(t_i-t_j)$ is generated and properly normalized \cite{Rembold2014}
\begin{equation}
g^{(2)}(u,\tau) = \frac{TY}{N^2\Delta\tau\Delta u}\frac{N_{u,\tau}}{\Big(1-\frac{\tau}{T}\Big)\Big(1-\frac{|u|}{Y}\Big)} ~.
\label{eq14}
\end{equation}
Here, $T$ and $Y$ describe the acquisition time and spatial length and $\Delta \tau$, $\Delta u$ the histogram bin size. The factor $[(1-\tau/T)(1-|u|/Y)]^{-1}$ corrects $N_{u,\tau}$ for the finite acquisition time and spatial length. The correlation function is normalized such that $\langle g^{(2)}(u,\tau)\rangle = 1$.
\begin{figure}
\centering
\includegraphics[width=0.6\textwidth]{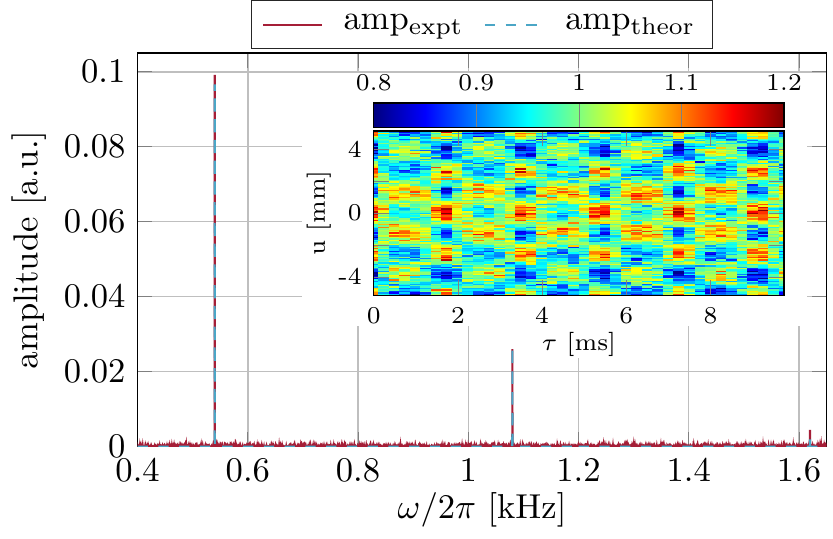} \caption{(Color online) Amplitude spectrum of the $\omega_0/2\pi = \unit[540]{Hz}$ measurement (solid red line), as calculated from the correlation function (see inset) via a discrete Fourier transformation at $u=N_u\lambda_{\text{g}^{(2)}}/2$. After identifying the fundamental perturbation frequency of $\omega_1/2\pi=\unit[(540.0\pm 0.05)]{Hz}$ equation (\ref{eq12}) is fitted to the spectrum (dashed blue line) and the peak phase deviation $\varphi_1=\unit[(0.5725\pm 0.0015)]{\pi}$ is obtained.}
\label{fig3}
\end{figure}

\begin{figure}
\centering
\includegraphics[width=0.6\textwidth]{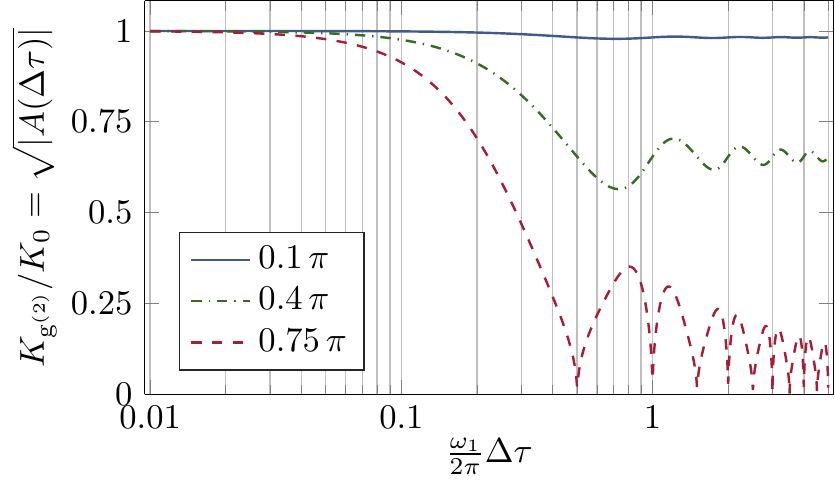} \caption{(Color online) Contrast reduction due to temporal binning for three different peak phase deviations \unit[0.1]{$\pi$}, \unit[0.4]{$\pi$} and \unit[0.75]{$\pi$} (blue solid line, green chain line and red dashed line). The curves are calculated according to Eq (\ref{eq15}).}
\label{fig4}
\end{figure}

The resulting correlation function for $\Delta u=$ \unit[90]{\textmu m} and $\Delta\tau=$ \unit[50]{\textmu s} is shown in the inset of figure \ref{fig3}. The spatial periodicity of the unperturbed interference pattern can be seen in $u$-direction. The periodicity in $\tau$-direction is $2\pi/\omega_1=\unit[1.9]{ms}$. The contrast of the unperturbed interference pattern is extracted at $\tau=0$ by fitting equation (\ref{eq11}) to the data. The results are $K_{\text{g}^{(2)}} = \unit[(58.5\pm 3.2)]{\%}$ and $\lambda_{\text{g}^{(2)}}= \unit[(2.60\pm0.02)]{mm}$. The extracted contrast, however, depends on the temporal binning $\Delta\tau$ of the correlation function. Following equation (\ref{eq9}), the bin averaged correlation function at $\tau=0$ and $N=1$ becomes
\begin{equation}
\frac{1}{\Delta\tau}\int_0^{\Delta\tau} g^{(2)}(u,\tau) \mathrm{d}\tau = 1 + \frac{1}{2}K_0^2\cdot A\left(\Delta\tau\right)\cos\left(ku\right) ~,
\label{eq15}
\end{equation}
with
\begin{equation}
A(\Delta\tau) = \sum_{m_1=-\infty}^\infty J_{m_1}\left(\varphi_1\right)^2\cdot \frac{\sin\left(m_1\omega_1\Delta\tau\right)}{m_1\omega_1\Delta\tau}~.\nonumber
\end{equation}
The extracted contrast $K_{\text{g}^{(2)}}$ is thus modified by the amplitude $A(\Delta\tau)$, reaching $K_0$ only in the limit $\Delta\tau\rightarrow 0$. Figure \ref{fig4} shows the expected contrast reduction $K_{\text{g}^{(2)}}/K_0=\sqrt{\left|A(\Delta\tau)\right|}$ due to temporal binning, for a single perturbation frequency $\omega_1$ and three different peak phase deviations $\varphi_j$ ranging from \unit[0.1]{$\pi$} to \unit[0.75]{$\pi$}. At these modulation strengths a binning of $\Delta\tau<0.08\cdot 2\pi/\omega_1$ is sufficient to extract the unperturbed contrast with \unit[95]{$\%$} accuracy. 

After having determined contrast and spatial periodicity of the unperturbed interference pattern, the perturbation frequency needs to be identified. This is done by calculating the temporal amplitude spectrum of the correlation function via a discrete Fourier transformation for every value $u=N_u\lambda_{\text{g}^{(2)}}/2$ with $N_u\in\mathbb{Z}$ and subsequent averaging. Following equation (\ref{eq12}), the resulting spectrum contains all frequency components $\omega_{\{m_1\}}$, as can be seen in figure \ref{fig3} for the \unit[540]{Hz} measurement (red solid line). Three peaks can be identified, that correspond to the fundamental frequency and harmonics at discrete frequencies $m_1\omega_1, m_1\in\{1,2,3\}$ in equation (\ref{eq12}). The amplitudes are given by the Bessel functions of the peak phase deviation and the contrast of the unperturbed interference pattern $\frac{1}{2}K_0^2J_{m_1}(\varphi_1)^2$. 
\begin{figure*}
\centering
\includegraphics[width=1\textwidth]{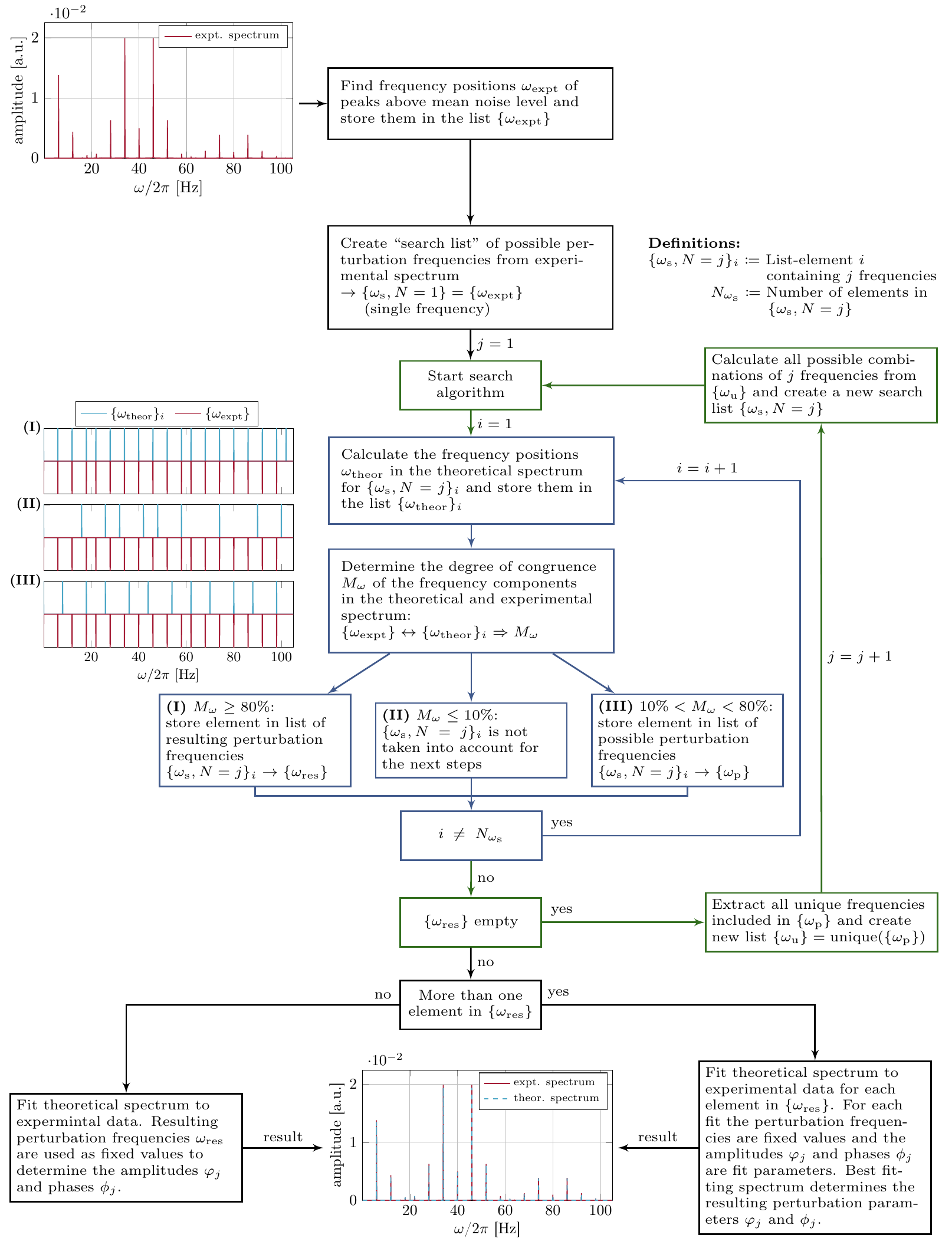} \caption{(Color online) Search algorithm for the identification of perturbation frequencies using the temporal amplitude spectrum of the correlation function.}
\label{fig5}
\end{figure*} 

For more than one perturbation frequency $\omega_j$, the spectrum consists not only of the fundamental frequencies and harmonics, but also of intermodulation terms. Therefore, it is difficult to identify the fundamental frequencies. For the determination of the correct perturbation frequencies a search algorithm has been developed which is described below and illustrated in figure \ref{fig5}. The algorithm will be discussed exemplary on a two-frequency perturbation with $\omega_1/2\pi=\unit[6]{Hz}, \varphi_1=\unit[0.6]{\pi}, \phi_1=\unit[0]{\pi}$ and $\omega_2/2\pi=\unit[40]{Hz}, \varphi_2=\unit[0.5]{\pi}, \phi_2=\unit[0]{\pi}$. This results in an amplitude spectrum of the correlation function as shown in figure \ref{fig5} top left (red solid line). First, the frequency positions $\omega_{\text{expt}}$ of all peaks above the mean noise level in the experimental amplitude spectrum are identified and stored in a list $\{\omega_{\text{expt}}\}$. For the example shown in figure \ref{fig5}, this list is $\{\omega_{\text{expt}}\}/2\pi=\{6,12,18,22,\ldots,98\}$. The next step is the creation of a ``search list" of possible perturbation frequencies from the experimental spectrum $\{\omega_{\text{s},N=1}\}=\{\omega_{\text{expt}}\}$ (single frequency case). Here, $\{\omega_{\text{s},N=j}\}_i$ is the list-element $i$ containing $j$ frequencies. The total number of elements in $\{\omega_{\text{s},N=j}\}$ is $N_{\omega_\text{s}}$. The search algorithm starts with the single frequency case ($j=1$) and uses the first element $i=1$ of $\{\omega_{\text{s},N=1}\}$. In the example, this is $\{\omega_{\text{s},N=1}\}_1/2\pi = \unit[6]{Hz}$. It is used as perturbation frequency $\omega_1$ in equation (\ref{eq12}) to calculate the positions of the frequency components $\omega_{\text{theor}}$ in the theoretical amplitude spectrum. The resulting positions are stored in the list $\{\omega_{\text{theor}}\}_i$ with $i=1$. For the perturbation frequency of \unit[6]{Hz} this list is $\{\omega_{\text{theor}}\}_1/2\pi=\{6,12,18,24,\ldots,96\}$. By comparing the frequency components in the theoretical and experimental spectrum, $\{\omega_{\text{theor}}\}_i$ and $\{\omega_{\text{expt}}\}$, the degree of congruence $M_\omega$ is determined. This indicates how many frequencies in the experimental spectrum coincide with those in the theoretical, compared to $N_{\omega_\text{expt}}$, the total number of frequencies stored in $\{\omega_{\text{expt}}\}$. In the example, three frequency positions coincide, $\{6,12,18\}$, and $N_{\omega_\text{expt}}=17$. This results in a degree of congruence of $M_\omega=\unit[17.6]{\%}$. The value of $M_\omega$ determines which of the next three cases is fulfilled. If $M_\omega\geq\unit[80]{\%}$, the element $\{\omega_{\text{s},N=1}\}_1$ is stored in the list of resulting perturbation frequencies $\{\omega_{\text{res}}\}$. The element $\{\omega_{\text{s},N=1}\}_1$ is not taken into account for the next steps, if $M_\omega\leq\unit[10]{\%}$. For $\unit[10]{\%}<M_\omega<\unit[80]{\%}$, the element is stored in a list of possible perturbation frequencies $\{\omega_{\text{p}}\}$. The third case is fulfilled for the element $\{\omega_{\text{s},N=1}\}_1/2\pi = \unit[6]{Hz}$. Then, the next element ($i=i+1$) in the list $\{\omega_{\text{s},N=1}\}$, $\{\omega_{\text{s},N=1}\}_2/2\pi = \unit[12]{Hz}$, is taken for the calculation of $\{\omega_{\text{theor}}\}_i$ and the determination of the degree of congruence $M_\omega$. This loop (indicated by the blue arrows and boxes in figure \ref{fig5}) continues until the last element in $\{\omega_{\text{s},N=1}\}$ was used ($i=N_{\omega_\text{s}}$). For the example in figure \ref{fig5}, the list of possible perturbation frequencies is $\{\omega_{\text{p}}\}/2\pi=\{6,12,18,24,\ldots,92\}$. The last element in $\{\omega_{\text{s},N=1}\}$, \unit[98]{Hz}, is missing because it fulfilled the second case. If no element has satisfied the first case ($M_\omega\geq\unit[80]{\%}$), then $\{\omega_{\text{res}}\}$ is empty and a new list $\{\omega_{\text{u}}\}$ is created containing all unique frequencies included in $\{\omega_{\text{p}}\}$. In the single frequency case it is trivial because $\{\omega_{\text{u}}\}=\{\omega_{\text{p}}\}$. As example, for two frequencies per element in $\{\omega_{\text{p}}\}/2\pi=\{\{7,13\},\{7,19\},\{13,41\}\}$ the list of unique frequencies would be $\{\omega_{\text{u}}\}/2\pi=\{7,13,19,41\}$. Using $\{\omega_{\text{u}}\}$, all possible combinations of $j=j+1$ frequencies are calculated and a new search list $\{\omega_{\text{s},N=2}\}/2\pi=\{\{6,12\},\{6,18\},\{6,22\},\ldots,\{86,92\}\}$ is created. Afterwards, the search algorithm starts  with the first element ($i=1$) in $\{\omega_{\text{s},N=2}\}$. This loop which increases the number of frequencies ($j=j+1$) is indicated in figure \ref{fig5} with green arrows and boxes. Then, all elements $i$ in $\{\omega_{\text{s},N=2}\}$ are probed by the search algorithm in the same way as described above for the single frequency case. On the left side of figure \ref{fig5}, the three cases are illustrated with different lists of frequency components $\{\omega_{\text{theor}}\}_i$ (blue solid line) together with $\{\omega_{\text{expt}}\}$ (red solid line). The first case was calculated for $\{\omega_{\text{s},N=2}\}_i/2\pi=\{\unit[6]{Hz},\unit[40]{Hz}\}$, the second for $\{\omega_{\text{s},N=2}\}_i/2\pi=\{\unit[58]{Hz},\unit[74]{Hz}\}$ and the third for $\{\omega_{\text{s},N=2}\}_i/2\pi=\{\unit[18]{Hz},\unit[62]{Hz}\}$. A very good match between theory and experiment can be seen for the first case. If the algorithm has stored one element in $\{\omega_{\text{res}}\}$ it is used as perturbation frequencies for the fit of equation (\ref{eq12}) to the experimental spectrum. Thereby, the extracted contrast $K_{\text{g}^{(2)}}$, spatial periodicity $\lambda_{\text{g}^{(2)}}$ and the perturbation frequencies $\omega_j$ are fixed parameters for the determination of the peak phase deviation $\varphi_j$ and phases $\phi_j$. If the list of resulting perturbation frequencies $\{\omega_{\text{res}}\}$ contains more than one element, a theoretical spectrum is fitted to the experimental data with each element of $\{\omega_{\text{res}}\}$ in the same way as described above for one element. The best matching theoretical spectrum determines the perturbation frequencies $\omega_j$ and parameters $\varphi_j$ and $\phi_j$. At the bottom of figure \ref{fig5} the theoretical spectrum (dashed blue line) for the resulting perturbation frequencies $\omega_1/2\pi=\unit[6]{Hz}$ and $\omega_2/2\pi=\unit[40]{Hz}$ can be seen. The algorithm identifies the perturbation frequencies $\omega_j$ with a probability of about \unit[90]{$\%$} for $N=3$ and \unit[70]{$\%$} for $N=4$.

With the exemplary measurement from figure \ref{fig2} and \ref{fig3}, the above described algorithm yields a single perturbation frequency $\omega_1/2\pi=\unit[(540.0\pm 0.05)]{Hz}$. After identification of the perturbation frequencies, equation (\ref{eq12}) is used to fit the peak phase deviation $\varphi_j$ to the amplitude spectrum, as shown in figure \ref{fig3} (dashed blue line). The resulting peak phase deviation is $\varphi_1=\unit[(0.5725\pm 0.0015)]{\pi}$. The main error in the determination of the perturbation parameters $\omega_j, \varphi_j$ and $\phi_j$ originates from the frequency resolution in the numerical amplitude spectrum. This was set to \unit[100]{mHz} to reduce the computing time, especially for the search algorithm.

With the obtained values from the correlation analysis, it is possible to reconstruct the interference pattern from the perturbed one in figure \ref{fig2}(a). To get the reconstructed pattern the new particle coordinates $y_{i,new}$ have to be calculated according to the extracted perturbation frequencies $\omega_j$ and peak phase deviations $\varphi_j$ \cite{Guenther2015}
\begin{equation}
y_{i,new} = y_i - \frac{\lambda}{2\pi}\varphi(t_i) = y_i - \frac{\lambda}{2\pi}\sum_{j=1}^N \varphi_j \cos\left(\omega_j t_i + \phi_j\right)~, 
\label{eq16}
\end{equation} 
where $\varphi(t_i)$ is the time-dependent phase shift in equation (\ref{eq2}) and $y_i$, $t_i$ are the spatial and temporal particle coordinates of the perturbed interference pattern. To determine the contrast of the reconstructed pattern, a two-dimensional histogram is calculated with $y_{i,new}$, averaged along the $x$-direction and fitted using equation (\ref{eq13}). To maximize the resulting contrast the perturbation frequencies $\omega_j$, the peak phase deviations $\varphi_j$ and phases $\phi_j$ are optimized by varying their values in a narrow window around the values extracted from the correlation analysis. For the exemplary measurement, this optimization results in $\omega_1/2\pi = \unit[539.994]{Hz}$, $\varphi_1 = \unit[0.66]{\pi}$ and $\phi_1 = \unit[0.59]{\pi}$. The reconstructed interference pattern, which can be seen in figure \ref{fig2}(b), reveals a contrast $K_{\text{rec}}=\unit[(55.5\pm 2.8)]{\%}$ and pattern periodicity $\lambda_{\text{rec}}=\unit[(2.57\pm0.01)]{mm}$. Both are determined similarly as for the perturbed interference pattern with equation (\ref{eq13}). The result for the reconstructed contrast is in good agreement with the contrast obtained from the correlation analysis $K_{\text{g}^{(2)}} = \unit[(58.5\pm 3.2)]{\%}$. 

The contrast in the reconstructed interference pattern depends strongly on the extracted perturbation values. Only, if the exact values of the perturbation ($\omega_j$, $\varphi_j$ and $\phi_j$) are used in equation (\ref{eq16}), the reconstructed contrast $K_{\text{rec}}$ is equal to the one of the unperturbed interference pattern $K_0$. For values with a deviation from the exact values $K_{\text{rec}}(|\Delta\omega|,\Delta\varphi,\Delta\phi)$ is reduced.
\begin{figure}
\centering
\includegraphics[width=0.6\textwidth]{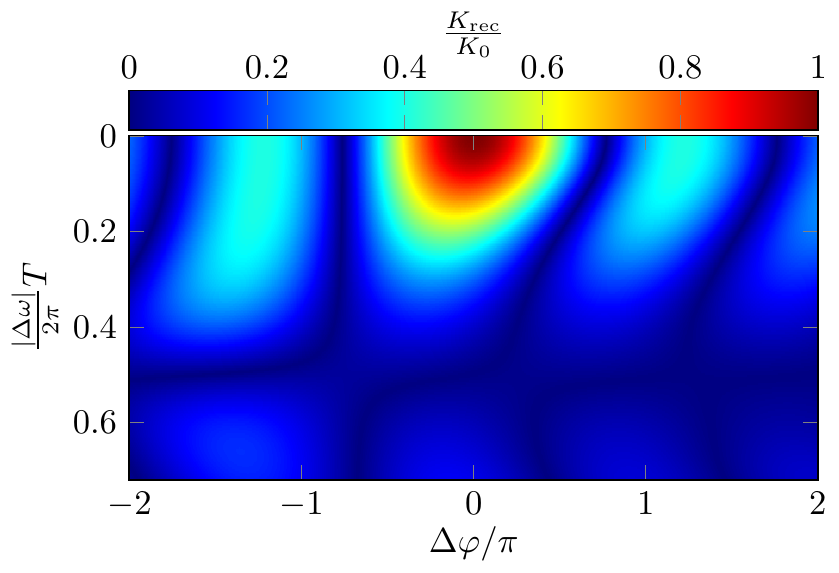} \caption{(Color online) Resulting contrast reduction $K_{\text{rec}}(|\Delta\omega|,\Delta\varphi,\Delta\phi)/K_0$, as calculated for different deviations $\frac{|\Delta\omega|}{2\pi}T$ and $\Delta\varphi$ at $\Delta\phi=\unit[0]{\pi}$. The diagram applies for all integration times $T$.}
\label{fig6}
\end{figure}

In equation (\ref{eq13}) the model function yields the contrast $K_0$, if the reconstructed coordinates are equal to the coordinates of the unperturbed interference pattern $y_0$. The maximum contrast is obtained at positions, where $ky_0=2\pi M$ for $y_0=M\lambda$ with $M\in \mathbb{Z}$, resulting in $K_{\text{rec}}=K_0\cos(2\pi M)$. If not the exact perturbation parameters are used in equation (\ref{eq16}), the coordinates of the unperturbed interference pattern are not correctly determined and a phase factor $\Delta\tilde{\varphi}$ remains, that depends on the deviations $|\Delta\omega|$, $\Delta\varphi$ and $\Delta\phi$. This factor reduces the contrast of the reconstructed interference pattern $K_{\text{rec}}(|\Delta\omega|,\Delta\varphi,\Delta\phi)=K_0\cos(2\pi M+\Delta\tilde{\varphi})$. By integration over the acquisition time $T$ a theoretical description for one perturbation frequency can be found and reads
\begin{align} \label{eq17}
K_{\text{rec}}(|\Delta\omega|,\Delta\varphi,\Delta\phi) ~=~& \frac{K_0}{T}\int_0^T \cos\bigg(2\pi M+\varphi_1\cos\Big(\omega_1 t+\phi_1\Big)- \\ \nonumber
&-\left(\varphi_1+\Delta\varphi\right)\cos\Big(\left(\omega_1+|\Delta\omega|\right)t+\left(\phi_1+\Delta\phi\right)\Big)\bigg) \mathrm{d}t ~.
\end{align}
Figure \ref{fig6} shows the resulting contrast reduction $K_{\text{rec}}(|\Delta\omega|,\Delta\varphi,\Delta\phi)/K_0$ as function of the relevant parameters $\frac{|\Delta\omega|}{2\pi}T$ and $\Delta\varphi$ for $\Delta\phi=\unit[0]{\pi}$. The result is independent of $\omega_1$, $\phi_1$ and $T$ as long as $T\gg\frac{2\pi}{\omega_1}$, i.e. the measurement time is much larger than the cycle duration of the perturbation. For the three cases, where two of three deviations are equal to zero and one is small, approximate solutions can be found
\begin{align} \label{eq18}
K_{\text{rec}}(|\Delta\omega|,0,0) &\approx K_0\cdot \mbox{e}^{-\frac{1}{2}\left(\frac{\pi}{8}|\Delta\omega| T\varphi_1\right)^2} \approx K_0\left(1-\frac{1}{2}\left(\frac{\pi}{8}|\Delta\omega| T\varphi_1\right)^2\right) \\ \nonumber
K_{\text{rec}}(0,\Delta\varphi,0) &\approx K_0\cdot \big|\!J_0\left(\Delta\varphi\right)\big|\quad ~~~ \approx K_0\left|\left(1-\frac{\Delta\varphi^2}{4}\right)\right| \\ \nonumber
K_{\text{rec}}(0,0,\Delta\phi) &\approx K_0\cdot \big|\!J_0\left(\Delta\phi\cdot\varphi_1\right)\big|~ \approx K_0\left|\left(1-\frac{\Delta\phi^2}{4}\varphi_1^2\right)\right|~. 
\end{align}
Knowing the landscape of figure \ref{fig6}, it is possible to optimize the reconstruction of the interference pattern. At the position of  $K_{\text{rec}}=K_0$ also the values of the perturbation are correctly determined. Above theory has been demonstrated for one perturbation frequency, but can be applied also in the case of numerous frequencies, because in equation (\ref{eq16}) the perturbations are independent of each other and therefore can be recalculated successively. For each reconstruction step the resulting contrast gets larger, until it reaches $K_0$.

For the exemplary measurement with \unit[540]{Hz}, the acquisition time was $T=\unit[19.2]{s}$. Using equation (\ref{eq18}) with $\varphi_1 = \unit[0.66]{\pi}, \Delta\varphi=0, \Delta\phi=0$ and $\Delta\omega/2\pi=\unit[5]{mHz}$, the reconstructed contrast is reduced to $K_{\text{rec}}=0.88\cdot K_0$. For the reconstruction of the unperturbed interference pattern with equation (\ref{eq16}), a frequency accuracy of $\unit[\pm1]{mHz}$ is required for the optimization process to reveal the reconstructed contrast with less than $\unit[1]{\%}$ deviation from $K_0$.

\section{Results}\label{Results}
The following measurements will demonstrate the extraction of the unperturbed interference pattern contrast $K_0$ in the presence of dephasing. Additionally, we will determine the vibrational response spectrum of the interferometer, including the possibility to reconstruct the unperturbed interference pattern. 
\begin{figure*}
\centering
\includegraphics[width=1\textwidth]{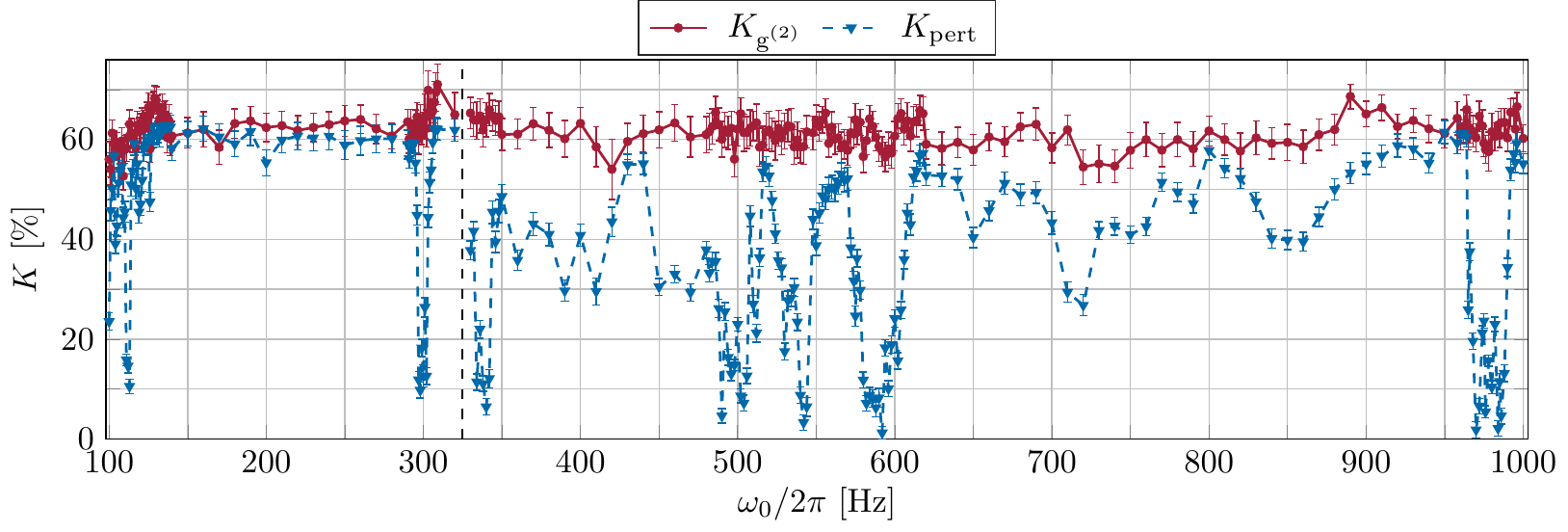} \caption{(Color online) Contrast $K_{\text{g}^{(2)}}$ (red dots with solid line) of the speaker and piezo measurements resulting from the correlation analysis using equation (\ref{eq11}) for $\tau = 0$ and the contrast of the perturbed interference pattern $K_{\text{pert}}$ (blue triangles with dashed line) obtained by fitting equation (\ref{eq13}) to the averaged histogram. The two measurement sets are separated by the black dashed vertical line at $\unit[325]{Hz}$. The averaged contrast $K_{\text{g}^{(2)}}$ over the complete measurement with the speaker is $\unit[(62.5\pm 3.4)]{\%}$ and the mean error of the individual fit is $\unit[\pm 3.0]{\%}$. For the piezo measurement the result is $\unit[(61.3\pm 2.7)]{\%}$ with the mean error $\unit[\pm 3.2]{\%}$. The averaged spatial periodicity for the speaker measurement is $\lambda_{\text{g}^{(2)}}= \unit[(2.62\pm0.04)]{mm}$ with the mean error of each fit $\unit[\pm 0.02]{mm}$ and $\lambda_{\text{pert}}= \unit[(2.60\pm0.04)]{mm}$ with the mean error of $\unit[\pm 0.05]{mm}$. For the piezo measurement the results are $\lambda_{\text{g}^{(2)}}= \unit[(2.65\pm0.05)]{mm}$ with $\unit[\pm 0.02]{mm}$ and $\lambda_{\text{pert}}= \unit[(2.65\pm0.04)]{mm}$ with $\unit[\pm 0.03]{mm}$.}
\label{fig7}
\end{figure*}

From the measurements with the speaker (excitation frequency $\omega_0/2\pi$ from 100 to \unit[320]{Hz}) and the piezo ($\omega_0/2\pi$ from 330 to \unit[1000]{Hz}) the correlation function is extracted according to equation (\ref{eq14}). For each measurement the correlation function is calculated with a spatial discretization of $\Delta u=$ \unit[90]{\textmu m} and a temporal of $\Delta \tau=$ \unit[200]{\textmu s} for the speaker and $\Delta \tau=$ \unit[50]{\textmu s} for the piezo measurement. The maximum correlation time is $\tau=\unit[10]{s}$. As discussed in section \ref{Data}, the contrast of the perturbed interference pattern $K_{\text{pert}}$ is determined by using equation (\ref{eq13}). From the correlation function at $\tau=0$ (equation (\ref{eq11})), the corresponding contrast of the unperturbed interference pattern $K_{\text{g}^{(2)}}$ is extracted. The results for the speaker and piezo measurement are shown in figure \ref{fig7}. The data for $K_{\text{pert}}$ show clear resonance structures at discrete frequencies. At these resonances, the contrast of the integrated interference image vanishes almost completely. However, the correlation analysis reveals the full contrast of the unperturbed interference pattern over the full spectral range. 
\begin{figure*}
\centering
\includegraphics[width=1\textwidth]{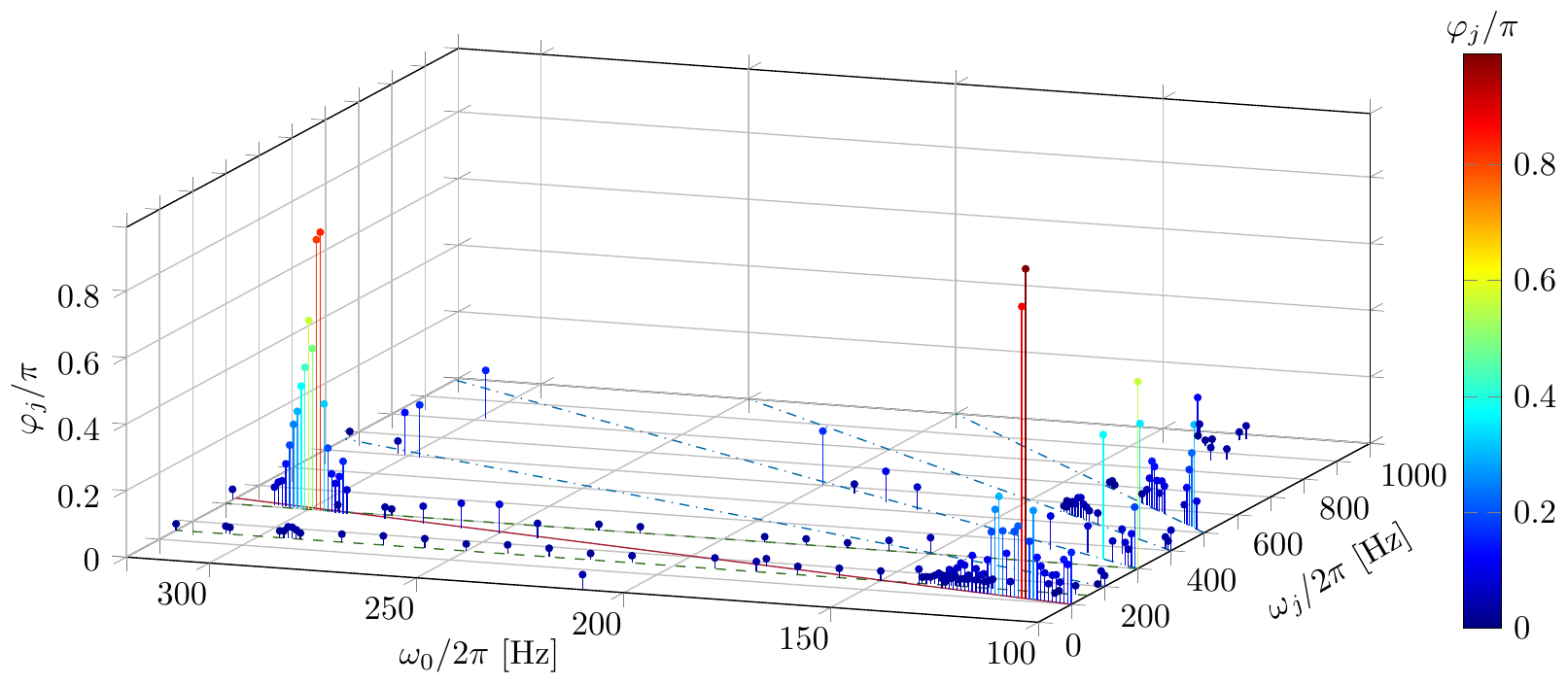} \caption{(Color online) Response spectrum of the interferometer as extracted from the speaker measurement. The red solid line in the frequency-plane shows the fundamental frequency, where $\omega_j$ is equal to $\omega_0$. The blue chain lines represent the higher harmonics of $\omega_0$. The green horizontal dashed lines indicate constant frequencies at \unit[150 and 300]{Hz}.}
\label{fig8}
\end{figure*}
\begin{figure*}
\centering
\includegraphics[width=1\textwidth]{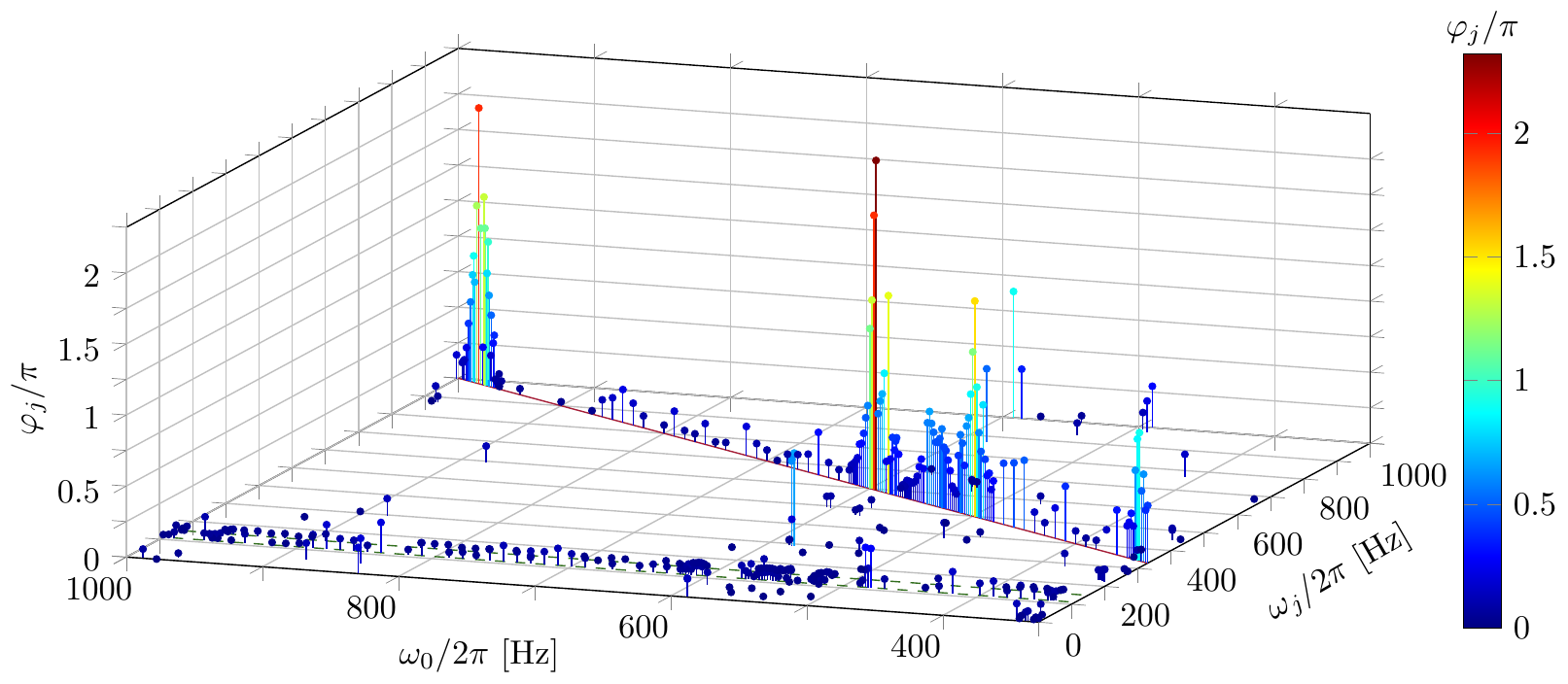} \caption{(Color online) Response spectrum of the piezo measurement. The red solid line shows the fundamental frequency of $\omega_0$. The green horizontal dashed lines are constant frequencies at \unit[111.4 and 150]{Hz}.}
\label{fig9}
\end{figure*}

Following section \ref{Data}, the amplitude spectrum of the correlation function is calculated and the involved perturbation frequencies $\omega_j$ and corresponding peak phase deviations $\varphi_j$ are identified by the described algorithm. The resulting response spectrum for the speaker measurement with excitation frequencies of $\omega_0/2\pi$ from 100 to \unit[320]{Hz} is shown in figure \ref{fig8}. The red solid line, plotted in the frequency-plane, denotes the positions of the fundamental frequency, where $\omega_j$ is equal to $\omega_0$. By comparison of the positions of large amplitudes $\varphi_j$ on this line with the positions of the reduced contrast $K_{\text{pert}}$ in figure \ref{fig7}, a good agreement can be seen, according to equation (\ref{eq3}). The maximum peak phase deviation in the complete spectrum is $\varphi_j=\unit[0.99]{\pi}$ at $\omega_j/2\pi=\unit[112]{Hz}$. The blue chain lines in the frequency-plane represent higher harmonics of $\omega_0$. For a given harmonic excitation at $\omega_0$ and a linear response of the interferometer, the response spectrum should include only the excitation frequency. However, due to anharmonicities in the excitation process and possible nonlinear response of the complex interferometer setup, the response spectrum may include higher harmonics. Especially in the region of 100 to $\unit[140]{Hz}$ this behaviour can be observed. The green horizontal dashed lines show constant frequencies at \unit[150 and 300]{Hz} that are independent of the excitation frequency, probably originating from the electrical network frequency at \unit[50]{Hz}.

The results of the piezo measurement with $\omega_0/2\pi$ from 330 to \unit[1000]{Hz} are shown in figure \ref{fig9}. Again, the positions of large peak phase deviations $\varphi_j$ agree well with the positions of reduced contrast $K_{\text{pert}}$ in figure \ref{fig7}, according to equation (\ref{eq3}). The maximum value in this spectrum is $\varphi_j=\unit[2.32]{\pi}$ at $\omega_j/2\pi=\unit[594]{Hz}$. The green horizontal dashed lines of constant frequencies are at \unit[111.4 and 150]{Hz}. The origin of the first one could be a vibration in the laboratory. The latter is likely a harmonic frequency of the electrical network. 

Comparing the two spectra in figure \ref{fig8} and \ref{fig9} it can be seen, that the resulting peak phase deviations of the speaker measurement are below the values extracted from the piezo measurement. One reason could be, that the excitation of the speaker is directed along the interferometer axis ($z$-direction, figure \ref{fig1}), whereas the piezo excitation is oriented along the direction of interference ($y$-direction). The piezo excitation will thus have a stronger influence on the dephasing of the interference pattern. Another difference between the two spectra are the positions of constant frequencies indicated by the green horizontal lines. This difference could probably originate from the excitation direction mentioned above and the fact that the two measurements have been made at different days. Therefore, a change of the environmental conditions could be possible.
\begin{figure}
\centering
\includegraphics[width=0.6\textwidth]{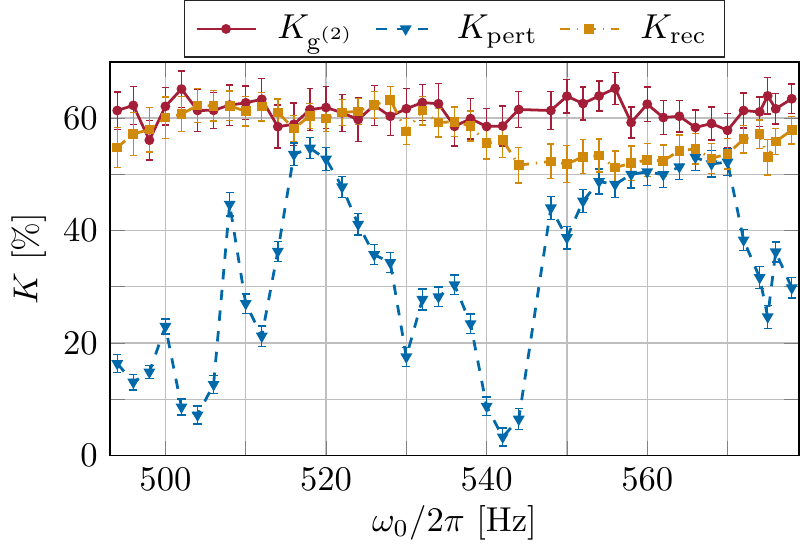} \caption{(Color online) Contrast obtained from the correlation analysis $K_{\text{g}^{(2)}}$ (red dots with solid line), contrast of the perturbed interference pattern $K_{\text{pert}}$ (blue triangles with dashed line) and contrast of the reconstructed interference pattern $K_{\text{rec}}$ (yellow squares with chain line) for different excitation frequencies $\omega_0$. To calculate $K_{\text{rec}}$ equation (\ref{eq16}) is used to maximize the contrast by varying the perturbation parameters. $K_{\text{g}^{(2)}}$ represents the upper limit for $K_{\text{rec}}$.}
\label{fig10}
\end{figure} 

Calculating the spatial periodicity of the electron interference pattern before the magnification through the quadrupole lens, allows to determine the spatial perturbation amplitude. The measurements have been made with an electron emission voltage of \unit[1.45]{kV} and a voltage of \unit[0.3]{V} at the biprism, yielding an interference pattern with the unmagnified spatial periodicity of $\lambda = \unit[880]{nm}$ \cite{Mollenstedt1956a}. With this value and the resulting peak phase deviations $\varphi_j$, the spatial perturbation amplitude $A$ can be calculated via $A(\varphi_j)=\lambda\frac{\varphi_j}{2\pi}$. The resulting amplitudes are in the range of \unit[6]{nm} ($\varphi_1=\unit[0.014]{\pi}$ at $\omega_1/2\pi = \unit[111.4]{Hz}$) up to \unit[1.021]{\textmu m} ($\varphi_1=\unit[2.320]{\pi}$ at $\omega_1/2\pi = \unit[594]{Hz}$). Here we assume, that the perturbation occurs before the magnification of the interference pattern through the quadrupole lens. In principle it is possible to increase the sensitivity for measuring perturbation amplitudes by reducing the spatial periodicity. This can be achieved by decreasing the acceleration voltage or increasing the biprism voltage.

With the obtained values from the correlation analysis, the unperturbed interference pattern can be reconstructed from the experimental data with equation (\ref{eq16}). In addition, the accuracy in identifying the perturbation frequencies $\omega_j$ and peak phase deviations $\varphi_j$ can be increased by maximising the contrast of the reconstructed pattern $K_{\text{rec}}$. The result is plotted in figure \ref{fig10} for the region from 494 to \unit[578]{Hz} of the piezo measurement. Over the whole range the contrast of the reconstructed interference pattern $K_{\text{rec}}$ is significantly larger than the contrast of the perturbed one $K_{\text{pert}}$ and close to the contrast $K_{\text{g}^{(2)}}$ of the unperturbed interference pattern, which marks an upper limit for the contrast of the reconstructed pattern.

\section{Conclusion}
Due to the technological progress for single-particle detection with high spatial and temporal resolution, it is possible to prove matter-wave interference by second-order correlation analysis, although the integrated interference structure vanishes by vibrational dephasing. Furthermore, the involved perturbation frequencies and amplitudes can be identified. In this paper we have demonstrated theoretically and experimentally, how this can be performed using the additional information about the particle impact time $t_i$ and position $y_i$. Our method can in principle be applied in various interferometric experiments equipped with such a detector. It has major potential for applications in sensor technology for vibrational as well as electromagnetic perturbations \cite{Rembold2014,Guenther2015}.

By introducing vibrations artificially to our biprism electron interferometer, we have disturbed the integrated interference pattern. The degree of disturbance depends on the response of the interferometer to this excitation frequency. With our method the matter-wave characteristics, contrast and spatial periodicity, were extracted for the whole excitation spectrum from 100 to \unit[1000]{Hz}. By calculating the amplitude spectrum of the correlation function, it is possible to identify the perturbation frequencies using our numerical search algorithm. With the obtained frequencies, the theoretical function was fitted to the experimental spectrum to determine the perturbation amplitudes and phases. By applying our method to all measurements, a response spectrum of the interferometer was created. With the possibility to reconstruct the unperturbed interference pattern, the accuracy of the parameters obtained from the correlation analysis could be increased.

We have demonstrated, that our method is not only a tool to prove matter-wave interferences even if they are invisible in the integrated picture, but has also possible sensor applications for the identification of perturbation frequencies and amplitudes originating from the environment. The interferometer could be used for the analysis of external perturbations, if its response spectrum is known. On the other hand, it is possible to apply a defined external perturbation source to map the response spectrum of the interferometer. With the knowledge of the vibrational as well as electromagnetic response spectrum of an interferometer, the mechanical decoupling from the environment and electromagnetic shielding can be optimized for the specific application. Additionally, perturbation sources can be identified and eliminated.

\section*{Acknowledgments}
This work was supported by the Deutsche Forschungsgemeinschaft (DFG, German Research Foundation) through the Emmy Noether program STI 615/1-1. A.G. acknowledges support from the DFG SFB TRR21, and A.R. from the Evangelisches Studienwerk e.V. Villigst. The authors thank N. Kerker and A. Pooch for helpful discussions.


\section*{References}
\begin{thebibliography}{01}

\bibitem{Grattan2013} Grattan L S and Meggitt B T 2013 eds.  \textit{Optical Fiber Sensor Technology: Fundamentals}, Springer Science \& Business Media 
\bibitem{Abbott2016} Abbott B P et al. 2016 \textit{Phys. Rev. Lett.} {\bf 116} 061102
\bibitem{Graham2013} Graham P W, Hogan J M, Kasevich M A and Rajendran S 2013 \textit{Phys. Rev. Lett} {\bf 110}, 171102
\bibitem{Gerlich2011} Gerlich S, Eibenberger S, Tomandl M, Nimmrichter S, Hornberger K, Fagan P J, T\"{u}xen J, Mayor M and Arndt M 2011 \textit{Nature Comm.} {\bf 2} 263
\bibitem{Arndt2014a} Arndt M and Hornberger K 2014 \textit{Nature Phys.} {\bf 10} 271
\bibitem{Haslinger2013} Haslinger P, D\"{o}rre N, Geyer P, Rodewald J, Nimmrichter S and Arndt M 2013 \textit{Nature Phys.} {\bf 9} 144  
\bibitem{gustavson1997} Gustavson T L, Bouyer P, and Kasevich M A 1997 \textit{Phys. Rev. Lett.} {\bf 78} 2046
\bibitem{Hasselbach1993} Hasselbach F and Nicklaus M 1993 \textit{Phys. Rev. A} {\bf 48} 143
\bibitem{Peters1999} Peters A, Chung K Y, and Chu S 1999 \textit{Nature} {\bf 400} 849
\bibitem{Margalit2015} Margalit Y, Zhou Z, Machluf S, Rohrlich D, Japha Y and Folman R 2015 \textit{Science} {\bf 349} 1205
\bibitem{Arndt2015a} Arndt M and Brand C 2015 \textit{Science} {\bf 349} 1168 
\bibitem{Stibor2005} Stibor A, Hornberger K, Hackerm\"{u}ller L, Zeilinger A and Arndt M 2005 \textit{Laser Physics} {\bf 15} 10
\bibitem{Miffre2006} Miffre A, Jacquey M,  B\"{u}chner M, Tr\'{e}nec G and Vigu\'{e} J 2006 \textit{Appl. Phys. B} {\bf 84} 617
\bibitem{Hauth2013} Hauth M, Freier C, Schkolnik V, Senger A, Schmidt M and Peters A 2013 \textit{Appl. Phys. B} {\bf 113} 49
\bibitem{Geiger2011} Geiger R, M\'{e}noret V, Stern G, Zahzam N, Cheinet P, Battelier B, Villing A, Moron F, Lours M, Bidel Y, Bresson A, Landragin A and Bouyer P 2011 \textit{Nature Commun.} {\bf 2} 474
\bibitem{Hensley1999} Hensley J M, Peters A and Chu S 1999 \textit{Rev. Sci. Instr.} {\bf 70} 2735
\bibitem{leGouet2008} Le Gou\"{e}t,
Mehlst\"{a}ubler T E, Kim J, Merlet S, Clairon A, Landragin A and Pereira dos Santos F 2008 \textit{Appl. Phys. B} {\bf 92} 133
\bibitem{Chiow2009} Chiow S W, Herrmann S, Chu S and M\"{u}ller H 2009 \textit{Phys. Rev. Lett.} {\bf 103} 050402
\bibitem{Chiow2011} Chiow S-W, Kovachy T, Chien H-C and Kasevich M A 2011 \textit{Phys. Rev. Lett} {\bf 107}, 130403
\bibitem{Fray2004} Fray S, Alvarez Diez C, H\"{a}nsch T W and Weitz M 2004 \textit{Phys. Rev. Lett.} {\bf 93} 240404
\bibitem{Chen2014} Chen X, Zhong J, Song H, Zhu L, Wang J and Zhan M 2014 \textit{Phys. Rev. A} {\bf 90} 023609
\bibitem{Barrett2015} Barrett B, Antoni-Micollier L, Chichet L, Battelier B, Gominet P-A, Bertoldi A, Bouyer P and Landragin A 2015 \textit{New J. Phys.} {\bf 17} 085010
\bibitem{Kohlhaas2015} Kohlhaas R, Bertoldi A, Cantin E, Aspect A, Landragin A and Bouyer P 2015 \textit{Phys. Rev. X} {\bf 5}, 021011
\bibitem{Hume2016} Hume D B and Leibrandt D R 2016 \textit{Phys. Rev. A} {\bf 93}, 032138
\bibitem{Zurek2003} Zurek W H 2003 \textit{Rev. Mod. Phys.} {\bf 75} 715
\bibitem{Pikovski2015} Pikovski I, Zych M, Costa F and Brukner \v{C} 2015 \textit{Nature Phys.} {\bf 11} 668
\bibitem{Hackermuller2004}	Hackerm\"{u}ller L, Hornberger K, Brezger B, Zeilinger A and Arndt M 2004 \textit{Nature} {\bf 427} 711
\bibitem{Hornberger2003} Hornberger K, Uttenthaler S, Brezger B, Hackerm\"{u}ller L, Arndt M and Zeilinger A 2003 \textit{Phys. Rev. Lett.} {\bf 90} 160401
\bibitem{Sonnentag2007} Sonnentag P and Hasselbach F 2007 \textit{Phys. Rev. Lett.} {\bf 98} 200402
\bibitem{Aharonov1959} Aharonov Y and Bohm D 1959 \textit{Phys. Rev.} {\bf 115} 485
\bibitem{Batelaan2009} Batelaan H and Tonomura A 2009  \textit{Physics Today} {\bf 62} 38
\bibitem{Schmid1985} Schmid H 1997 \textit{Dissertation, University of T\"{u}bingen}
\bibitem{Schuetz2015b} Sch\"{u}tz G, Rembold A, Pooch A, Prochel H and Stibor A 2015 \textit{Ultramicroscopy} {\bf 158} 65
\bibitem{Rembold2014} Rembold A, Sch\"{u}tz G, Chang W T, Stefanov A, Pooch A, Hwang I S, G\"{u}nther A and Stibor A 2014 \textit{Phys. Rev. A} {\bf 89} 033635
\bibitem{Guenther2015} G\"{u}nther A, Rembold A, Sch\"{u}tz G and Stibor A 2015 \textit{Phys. Rev. A} {\bf 92} 053607
\bibitem{Folling2005} F\"{o}lling S, Gerbier F, Widera A, Mandel O,
Gericke T and Bloch I 2005 \textit{Nature} {\bf 434} 481
\bibitem{Jagutzki2002} Jagutzki O, Mergel V, Ullmann-Pfleger K, Spielberger L, Spillmann U, D\"{o}rner R and Schmidt-B\"{o}cking H 2001 \textit{Nucl. Instr. Meth. Phys. Research A} {\bf 477} 244
\bibitem{Wiener1930} Wiener N 1930 \textit{Acta Mathematica} {\bf 55}(1) 117--258
\bibitem{Khintchine1934} Khintchine A 1934 \textit{Mathematische Annalen} {\bf 109}(1) 604--615
\bibitem{Siegmund2007} Siegmund O H, Vallerga J V, Tremsin A S, Mcphate J and Feller B 2007 \textit{Nucl. Instr. Meth. Phys. Research A} {\bf 576} 178
\bibitem{Schellekens2005} Schellekens M, Hoppeler R, Perrin A, Viana Gomes J, Boiron D, Aspect A and Westbrook C I 2005 \textit{Science} {\bf 310} 648
\bibitem{Zhou2012} Zhou X, Ranitovic P, Hogle C W, Eland J H D, Kapteyn H C, and Murnane M M 2012  \textit{Nature Physics} {\bf 8} 232
\bibitem{Mollenstedt1956a} M\"{o}llenstedt G and D\"{u}ker H 1956 \textit{Z. Phys. A - Hadron Nucl.} {\bf 145} 377
\bibitem{Schuetz2014} Sch\"{u}tz G, Rembold A, Pooch A, Meier S, Schneeweiss P, Rauschenbeutel A, G\"{u}nther A, Chang W T, Hwang I S and Stibor A 2014 \textit{Ultramicroscopy} {\bf 141} 9
\bibitem{Hasselbach2010} Hasselbach F 2010 \textit{Rep. Prog. Phys.} {\bf 73} 016101
\bibitem{Hasselbach1998a} Hasselbach F and Maier U 1999 \textit{Quantum Coherence and Decoherence - Proc. ISQM-Tokyo 98 ed. by Y.A. Ono and K. Fujikawa (Amsterdam: Elsevier)} 299
\bibitem{Maier1997} Maier U 1997 \textit{Dissertation, University of T\"{u}bingen}
\bibitem{Kuo2006a} Kuo H S, Hwang I S, Fu T Y, Lin Y C, Chang C C, and Tsong T T 2006 \textit{Japanese J. Appl. Phys.} {\bf 45} 8972
\bibitem{Kuo2008} Kuo H S, Hwang I S, Fu T Y, Lu Y H, Lin C Y and Tsong T T 2008 \textit{Appl. Phys. Lett.} {\bf 92} 063106
\end{thebibliography}
\end{document}